\documentclass[10pt,twoside,twocolumn,english,aps,manuscript,aps,preprint,show pacs,prb]{revtex4}
\usepackage[T1]{fontenc}
\usepackage[latin9]{inputenc}
\usepackage{babel}
\usepackage{array}
\usepackage{longtable}
\usepackage{textcomp}
\usepackage{amstext}
\usepackage{graphicx}
\usepackage{esint}
\usepackage[unicode=true,pdfusetitle,
 bookmarks=true,bookmarksnumbered=false,bookmarksopen=false,
 breaklinks=false,pdfborder={0 0 1},backref=false,colorlinks=false]
 {hyperref}
\usepackage{breakurl}

\makeatletter

\DeclareRobustCommand{\greektext}{%
  \fontencoding{LGR}\selectfont\def\encodingdefault{LGR}}
\DeclareRobustCommand{\textgreek}[1]{\leavevmode{\greektext #1}}
\DeclareFontEncoding{LGR}{}{}
\DeclareTextSymbol{\~}{LGR}{126}
\newcommand{\lyxmathsym}[1]{\ifmmode\begingroup\def\b@ld{bold}
  \text{\ifx\math@version\b@ld\bfseries\fi#1}\endgroup\else#1\fi}

\providecommand{\tabularnewline}{\\}

\@ifundefined{textcolor}{}
{%
 \definecolor{BLACK}{gray}{0}
 \definecolor{WHITE}{gray}{1}
 \definecolor{RED}{rgb}{1,0,0}
 \definecolor{GREEN}{rgb}{0,1,0}
 \definecolor{BLUE}{rgb}{0,0,1}
 \definecolor{CYAN}{cmyk}{1,0,0,0}
 \definecolor{MAGENTA}{cmyk}{0,1,0,0}
 \definecolor{YELLOW}{cmyk}{0,0,1,0}
}

\makeatother

\begin{document}

\title{A S=1/2 vanadium-based geometrically frustrated spinel system Li$_{\text{2}}$ZnV$_{\text{3}}$O$_{\text{8}}$ }

\author{T. Chakrabarty}

\email{tanmoyc@iitb.ac.in}

\affiliation{Department of Physics, IIT Bombay, Powai, Mumbai 400076, India }

\author{A. V. Mahajan}

\email{mahajan@phy.iitb.ac.in}

\affiliation{Department of Physics, IIT Bombay, Powai, Mumbai 400076, India}

\author{B. Koteswararao}

\affiliation{(CeNSCMR), Department of Physics and Astronomy, and Institute of
Applied Physics, Seoul National University, Seoul 151-747, South Korea}
\begin{abstract}
We report the synthesis and characterization of Li$_{2}$ZnV$_{3}$O$_{8}$,
which is a new Zn-doped LiV$_{2}$O$_{4}$ system containing only
tetravalent vanadium. A Curie-Weiss susceptibility with a Curie-Weiss
temperature of $\theta_{CW}\approx-214$ K suggests the presence of
strong antiferromagnetic correlations in this system. We have observed
a splitting between the zero-field cooled ZFC and field cooled FC
susceptibility curves below 6 K. A peak is present in the ZFC curve
around 3.5 K suggestive of spin-freezing . Similarly, a broad hump
is also seen in the inferred magnetic heat capacity around 9 K. The
consequent entropy change is only about 8\% of the value expected
for an ordered $S=1/2$ system. This reduction indicates continued
presence of large disorder in the system in spite of the large $\theta_{CW}$
, which might result from strong geometric frustration in the system.
We did not find any temperature $T$ dependence in our $^{7}$Li nuclear
magnetic resonance NMR shift down to 6 K (an abrupt change in the
shift takes place below 6 K) though considerable $T$-dependence has
been found in literature for LiV$_{2}$O$_{4}$$-$ undoped or with
other Zn/Ti contents. Consistent with the above observation, the $^{7}$Li
nuclear spin-lattice relaxation rate $1/T_{1}$ is relatively small
and nearly $T$-independent except a small increase close to the freezing
temperature, once again, small compared to undoped or 10\% Zn or 20\%
Ti-doped LiV$_{2}$O$_{4}$.
\end{abstract}


\maketitle

\section{introduction}

Cubic spinel materials AB$_{2}$O$_{4}$ with non-magnetic ions at
the A-site and magnetic ions at the B-site are interesting due to
the existence of a geometrically frustrated\cite{Leo Balents Nature,J.E.Greedan-Geometrically frustrated magnetic materials-2000}
B-sublattice (corner-shared tetrahedra) in them. Perhaps the most
studied compound in this category is LiV$_{2}$O$_{4}$ which exhibits
the $d$-electron derived heavy fermi liquid behaviour\cite{LiV2O4-HF behavior,Li2VO4-HF-2,Li2VO4-HF-3}.
In the motivation of Villain's work\cite{J Villain's paper} LiV$_{2}$O$_{4}$
doped with a non-magnetic impurity at the V-site or with a magnetic
impurity at the A-site have also been studied extensively\cite{Spin-glass behavior in Li1-xZnxV2O4,Anomalous structural behavior of Zn-doped LiV2O4,Li(V1-xMx)2O4(M=00003DCr Ti),Ground state properties od LiV2O4 and Li1-xZnx(V1-yTiy)2O4,Li NMR studies of Li1-xZnxV2O4 and Li(V1-yTiy)2O4}.
Likewise, the structure of Na$_{4}$Ir$_{3}$O$_{8}$ or {[} Na$_{1.5}$(Na$_{0.25}$Ir$_{0.75}$)$_{2}$O$_{4}${]}
is also derived from the spinel structure\cite{Na4Ir3O8}. Here, the
Na1.5 are at the A-sites whereas Ir and Na at the B-sites have distinct
positions and the Ir ions form corner shared triangles in three dimensions.
This has been dubbed as the hyperkagome lattice and is proposed to
have a quantum spin liquid ground state. 

Motivated by the above we have been searching for new $S=1/2$ spinel
systems which might exhibit interesting properties. There exists Li$_{2}$ZnTi$_{3}$O$_{8}$
which has the same space group as Na$_{4}$Ir$_{3}$O$_{8}$, i.e.,
\textit{P4$_{3}$32} and the Zn and the Ti ions are at distinct B-sites
therefore the Ti ions form a hyperkagome lattice. This compound is
unfortunately nonmagnetic due to the +4 oxidation state of the titanium.
We therefore set out to prepare Li$_{2}$ZnV$_{3}$O$_{8}$ which
would have a $S=1/2$ (V$^{4+}$) hyperkagome lattice, in the case
in which it formed with the same structure as Li$_{2}$ZnTi$_{3}$O$_{8}$.
In this paper we report the properties of a new V-based spinel Li$_{2}$ZnV$_{3}$O$_{8}$.
Our results show that Li$_{2}$ZnV$_{3}$O$_{8}$ does not form in
the \textit{P4$_{\text{3}}$32} space group and hence is not isostructural
to Li$_{2}$ZnTi$_{3}$O$_{8}$. In fact, Li$_{2}$ZnV$_{3}$O$_{8}$
forms in the \textit{F d -3 m-s} space group where there is no unique
B-site and it is shared by two ions. So, Li$_{2}$ZnV$_{3}$O$_{8}$
could be written as {[}Li(Zn$_{0.25}$V$_{0.75}$)$_{2}$O$_{4}${]}$_{2}$
or {[}(Li$_{0.5}$Zn$_{0.5}$(Li$_{0.25}$V$_{0.75}$)$_{2}$O$_{4}${]}$_{2}$
or possibly with a site occupation which is in between the two. In
the former site assignment, Li is at the tetrahedral A-site and the
Zn and V ions share the octahedral B-sites while in the latter, Li
and Zn are at the tetrahedral A-sites and Li and V ions share the
octahedral B-sites. Consequently, although the B-sublattice forms
a geometrically frustrated network, in both the cases only 75\% of
its sites are (statistically) occupied by the magnetic V$^{4+}$ ions.
Due to this site-sharing at the B-site, the frustration effect is
likely diluted and might result in a spin-disordered (frozen) state
at low temperature. In LiV$_{2}$O$_{4}$ the V$^{4+}$/V$^{3+}$
ratio is 1 and the system shows heavy fermionic behavior. When a non-magnetic
impurity (Zn,Ti) is doped at the B-site or at the A-site of the spinel
then the system exhibits spin freezing or spin glass behaviour at
a temperature which depends on the doping element and its concentration\cite{Spin-glass behavior in Li1-xZnxV2O4,Li(V1-xMx)2O4(M=00003DCr Ti),magnetic and structural transition in LixZn1-xV2O4}.
In all these above mentioned doped LiV$_{2}$O$_{4}$ systems there
is likely a mixture of V$^{4+}$ and V$^{3+}$ions. In contrast, in
Li$_{2}$ZnV$_{3}$O$_{8}$ the effective valence of vanadium is 4,
which to our knowledge is the first doped-LiV$_{2}$O$_{4}$ system
but still with only V$^{4+}$($S=1/2$) ions.

The magnetic susceptibility of Li$_{2}$ZnV$_{3}$O$_{8}$ is found
to be of the Curie-Weiss form above 150 K with an effective moment
close to that expected for $S$ = $1/2$. The Curie-Weiss temperature
$\theta_{CW}$ = -214 K is indicative of strong antiferromagnetic
(AF) interactions. This value of the Curie-Weiss temperature is quite
high compared to the earlier doped LiV$_{2}$O$_{4}$ systems indicating
much stronger correlation between the magnetic B-sites of the system.
For the series Li$_{1-x}$Zn$_{x}$V$_{2}$O$_{4}$ the maximum value
of $\theta_{CW}$ is about -68 K (for $x=0.3$)\cite{Ground state properties od LiV2O4 and Li1-xZnx(V1-yTiy)2O4}
and for the series Li(V$_{1-y}$Ti$_{y}$)$_{2}$O$_{4}$ the maximum
value of $\theta_{CW}$ is about -56 K (for $y=0.3$)\cite{Ground state properties od LiV2O4 and Li1-xZnx(V1-yTiy)2O4}.
A difference between the zero-field cooled (ZFC) and field-cooled
(FC) susceptibilities is observed by us for Li$_{2}$ZnV$_{3}$O$_{8}$,
below about $6$ K, suggesting random magnetic interactions and spin-freezing.
A magnetic contribution to the heat capacity is present at low temperatures
though with a small entropy change, suggesting a highly degenerate
ground state. We also report $^{7}$Li NMR spectra and spin-lattice
relaxation rate ($1/T_{1}$) measurements and compare these with the
results for pure and other Zn/Ti doped LiV$_{2}$O$_{4}$.

\section{experimental details}

\begin{figure}
\begin{centering}
\includegraphics[scale=0.3]{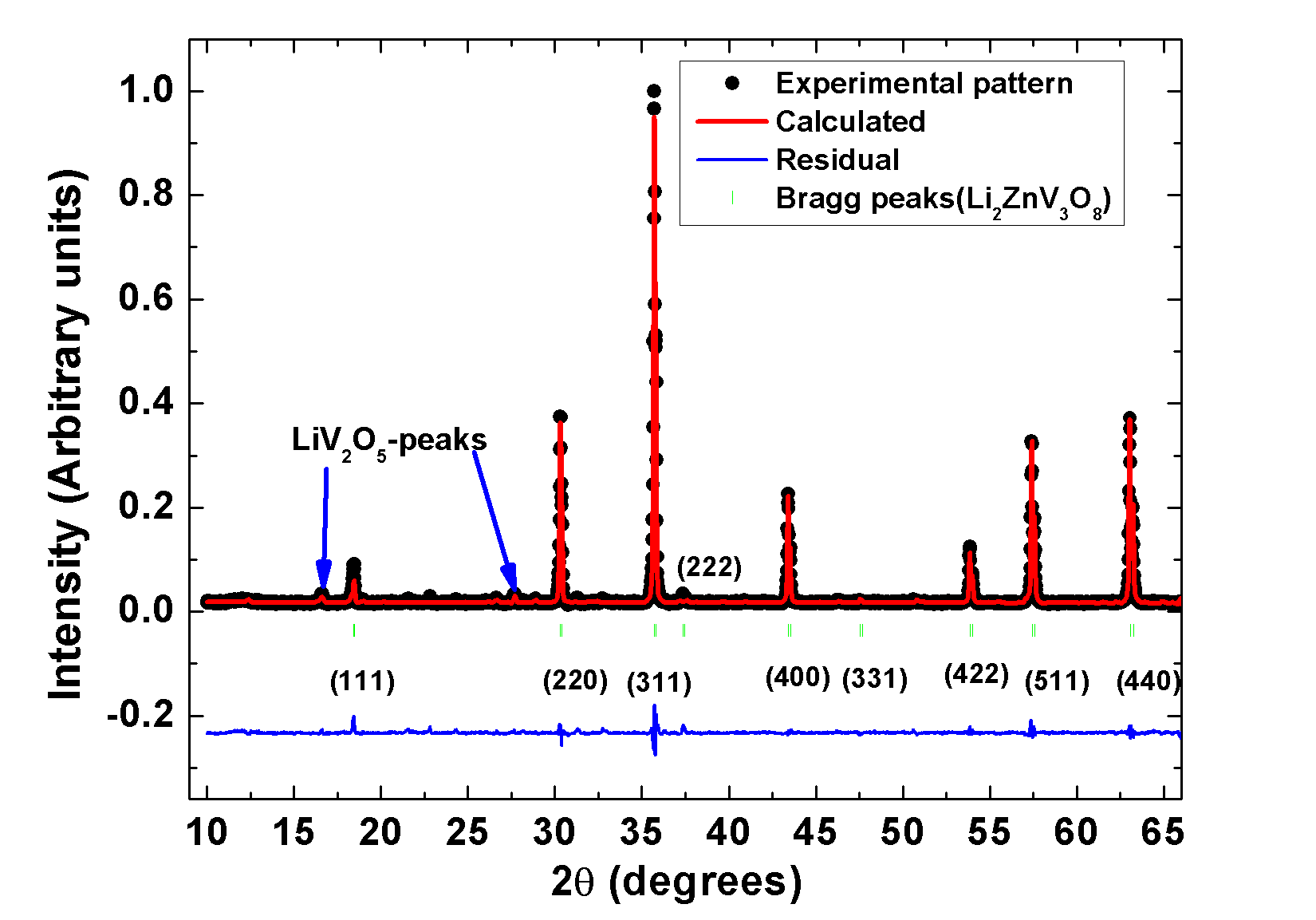}
\par\end{centering}

\caption{\label{fig:1xrd}Powder diffraction pattern of Li$_{2}$ZnV$_{3}$O$_{8}$
is shown along with its Bragg peak positions; The black points are
the experimental data, the red solid line is the ``Fullprof'' generated
refinement pattern, the green markers are the Bragg peak positions
for\textit{ F d -3 m-s} space group and the blue points represent
the ``experimental - calculated'' intensity pattern. LiV$_{2}$O$_{5}$
impurity peaks are shown by blue arrows. }
\end{figure}

Li$_{2}$ZnV$_{3}$O$_{8}$ was prepared by standard solid state reaction
methods. First we prepared V$_{2}$O$_{3}$ by reducing V$_{2}$O$_{5}$
(Aldrich\textemdash{}99.99\%) in hydrogen atmosphere at $650^{\circ}$C
for 16 hours. Then we prepared VO$_{2}$ by mixing this V$_{2}$O$_{3}$with
V$_{2}$O$_{5}$ in a 1:1 molar ratio, pelletizing and firing in dynamical
vacuum (better than $10^{-5}$ mbar) at $800^{\circ}$C for 24 hours.
In the next step we mixed Li$_{2}$CO$_{3}$ (Alpha Aesar 99.995\%
purity), ZnO (Aldrich 99.99\% purity) and VO$_{2}$, pelletized it,
and then fired in a tubular furnace at $650^{\circ}$C for 28 hours
under a dynamical vacuum (better than 10$^{-5}$ mbar). X-ray diffraction
(xrd) patterns were collected with a PANalytical x-ray diffractometer
using Cu K$\alpha$ radiation ($\lambda=\mbox{1.54182 }\textrm{\ensuremath{\mathrm{\AA}}}$).
The Rietveld refinement of the xrd pattern (see Fig.\ref{fig:1xrd})
was carried out using the ``Fullprof''\cite{Full prof} software.
The lattice parameter was found to be 8.332 Å in the \textit{F d -3
m} space group. The formula of Li$_{2}$ZnV$_{3}$O$_{8}$ could be
written in two different ways, {[}Li(Zn$_{0.25}$V$_{0.75}$)$_{2}$O$_{4}${]}$_{2}$
and {[}(Li$_{0.5}$Zn$_{0.5}$(Li$_{0.25}$V$_{0.75}$)$_{2}$O$_{4}${]}$_{2}$.
Out of the two possible site arrangements in Li$_{\text{2}}$ZnV$_{\text{3}}$O$_{\text{8}}$,
the second one is possibly the right site assignment since Zn is known
to prefer the tetrahedral site in spinels.\cite{V3=00003D-V4+ in mixed vanadium oxides LiZnxV2-xO4 and LiMgxV2-xO4}.
In both these possibilities V is at the B-site. We refined the xrd
data of Li$_{\text{2}}$ZnV$_{\text{3}}$O$_{\text{8}}$ to determine
the site occupation. The best refinement was obtained with the formula
{[}(Li$_{0.46}$Zn$_{0.43}$)(Li$_{0.24}$V$_{0.75}$)$_{2}$O$_{4}${]}$_{2}$
which imples some Zn-deficiency at the A-site. The results of the
refinements are shown in Table \ref{atomic positions} below. The
goodness of the Rietveld refinement is defined by the following parameters.
R$_{p}$= 19.6\%, R$_{wp}$ = 16.9\%, R$_{exp}$= 5.30\%, and \textgreek{q}$^{2}$
= 11.3. There were some extra phases of LiV$_{2}$O$_{5}$, Li$_{3}$VO$_{4}$
and VO$_{2}$ present at the 1\% level. The Rietveld refinement is
done considering all the 4 phases. 

From ICP-AES (inductively coupled plasma-atomic emission spectroscopy)
we determined the ratio of lithium with zinc and vanadium. From this
analysis we found the molar formula unit of the compound to be Li$_{1.98}$Zn$_{0.86}$V$_{3}$O$_{8}$.This
result is consistent with our ``Fullprof'' refinement since there
also we found that zinc is deficient at the A-site.

\begin{table}
\caption{\label{atomic positions}Atomic positions in Li$_{2}$ZnV$_{3}$O$_{8}$}

\begin{longtable}{|c|c|c|c|c|}
\hline 
Atoms & \multicolumn{1}{c}{} & \multicolumn{1}{c}{Co-ordinates} &  & Occupancy\tabularnewline
\cline{2-5} 
 & x/a & y/b & z/c & \tabularnewline
\hline 
Li1(8a) & 0.000 & 0.000 & 0.000 & 0.46\tabularnewline
\hline 
Zn1(8a) & 0.000 & 0.000 & 0.000 & 0.43\tabularnewline
\hline 
Li2(16d) & 0.625 & 0.625 & 0.625 & 0.24\tabularnewline
\hline 
V(16d) & 0.625 & 0.625 & 0.625 & 0.75\tabularnewline
\hline 
O1(32e) & 0.594 & 0.594 & 0.594 & 1.00\tabularnewline
\hline 
\end{longtable}
\end{table}

\begin{figure}
\includegraphics[scale=0.45]{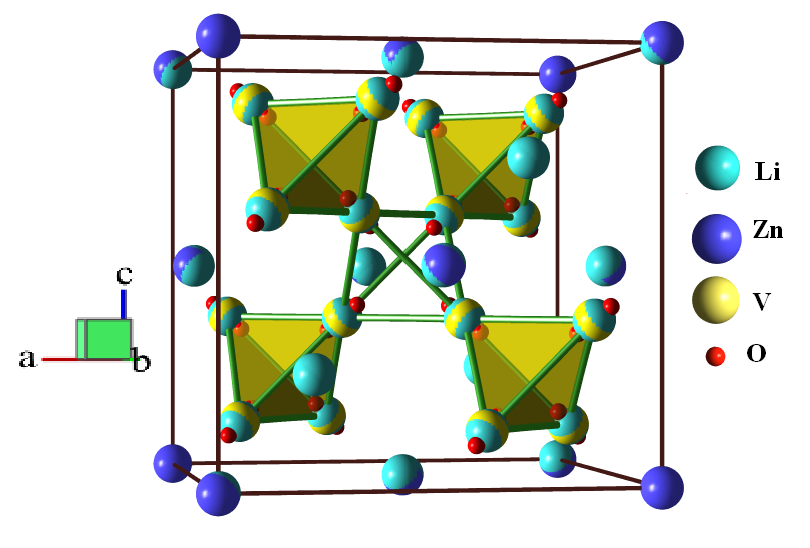}

\caption{\label{fig:2unit cell}Unit cell of Li$_{2}$ZnV$_{3}$O$_{8}$. The
Li$^{+}$, Zn$^{2+}$,V$^{4+}$,O$^{2-}$ions are shown in indigo,
blue, yellow and red color respectively. The B-sites form a corner-shared
tetrahedral network but are shared by Li$^{+}$/Zn$^{2+}$ and V$^{4+}$. }
\end{figure}

In the structure of Li$_{2}$ZnV$_{3}$O$_{8}$ the B-sites (shared
by Zn$^{2+}$and V$^{4+}$ in the 1:3 ratio) form a corner-shared
tetrahedral network {[}see Fig.\ref{fig:2unit cell}{]}. In case of
random/statistical occupation of the B sites by Zn and V, there are
going to be missing magnetic atoms in the triangular network. This
disruption/dilution of the corner-shared tetrahedral network is likely
to lead to relieving of frustration and this might then lead to a
spin-glass like state at low temperature. 

The temperature dependence of magnetization $M$ was measured in a
1 kOe magnetic field $H$ in the temperature range 2-300 K using a
vibrating sample magnetometer (VSM) attached with a Quantum Design
Physical Property Measurement System (PPMS). The temperature dependence
of heat capacity has also been measured in the temperature range of
2-270 K using the heat capacity attachment of a Quantum Design PPMS.
The $^{7}$Li (gyromagnetic ratio $\gamma/2\pi$ = 16.54607 MHz/kOe
and nuclear spin $I=3/2$) NMR measurements using a fixed field of
$93.9543$ kOe have been carried out. The \textsuperscript{7}Li nuclear
spin-lattice relaxation was measured by the saturation recovery method
using a $n\times\lyxmathsym{\textgreek{p}}/2$ - $t$ - $\lyxmathsym{\textgreek{p}}/2$
- $\tau$ - $\lyxmathsym{\textgreek{p}}$ pulse sequence with $\tau$
= $90$ $\lyxmathsym{\textgreek{m}}s$ and variable $t$. A comb sequence
with $n$ (between 5 and 10) pulses was used to obtain good saturation.

\section{results and discussions}

\begin{figure}
\includegraphics[scale=0.32]{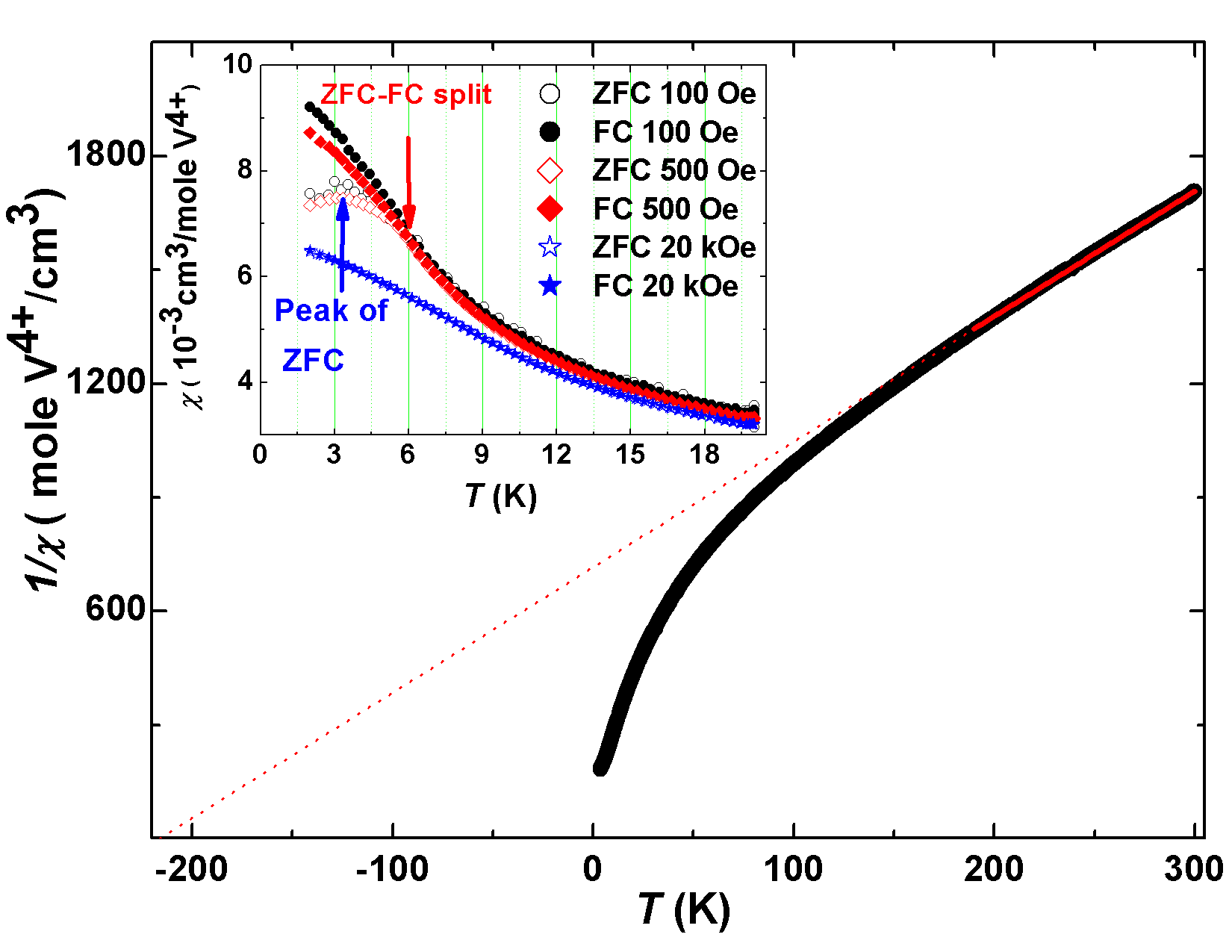}

\caption{\label{fig:3ZFC-FC}Temperature dependence of inverse susceptibility
$1/(\chi-\chi_{0})$ is shown for $H=10$ kOe. The red line shows
the Curie-Weiss fit in the temperature range of 190-300 K and the
dotted line is its extrapolation.The inset shows the bifurcation between
the ZFC and FC curves below about 6 K. The ZFC and FC data are shown
for various fields. The peak of the ZFC curve and the starting point
of ZFC/FC bifurcation have been pointed out by the blue and red arrow,
respectively.}
\end{figure}

\begin{figure}
\includegraphics[scale=0.35]{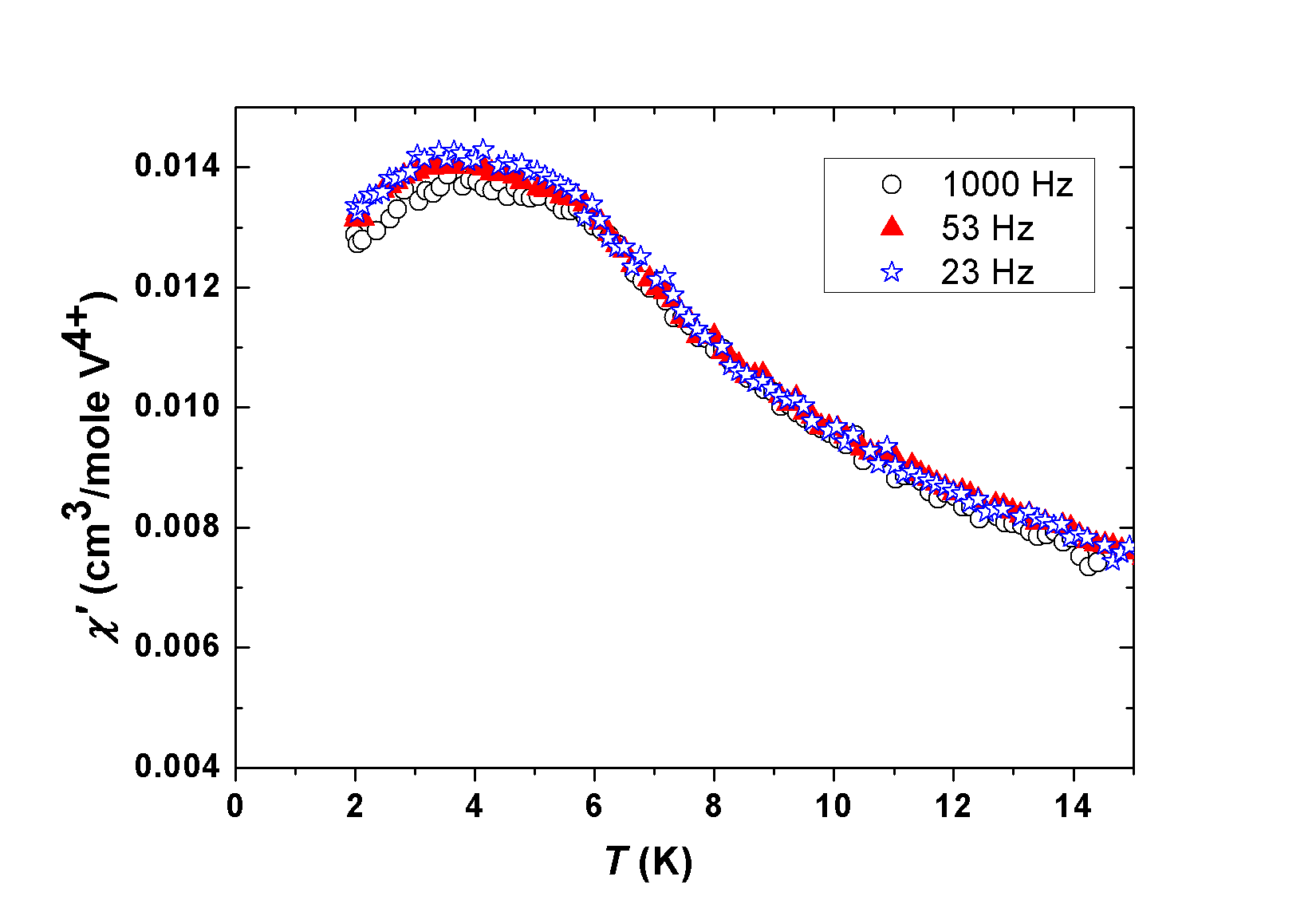}\caption{\label{fig:4ac susceptibility}Temperature dependence of the $ac$
susceptibility ($\chi'$) measured in an $ac$ field of 1 Oe. The
black, red and blue data points represent the measurements at 1000
Hz, 53 Hz, and 23 Hz respectively. A hump-like anomaly is observed
at 4 K in $\chi'$ but no dependence on frequency has been observed. }
\end{figure}

The reciprocal of the susceptibility is plotted with temperature in
Fig. \ref{fig:3ZFC-FC}. We fit the data to $1/(\lyxmathsym{\textgreek{q}}-\lyxmathsym{\textgreek{q}}_{\text{0}})=(T-$$\theta_{CW}$)$/C$
in the range 190 - 300 K, where we took $\lyxmathsym{\textgreek{q}}_{\text{0}}$
as the sum of the core diamagnetic susceptibilites for all the atoms
in Li$_{2}$ZnV$_{3}$O$_{8}$ \cite{Core diamagnetic susceptibiilty}
and the Van-Vleck paramagnetic susceptibility $\chi_{VV}=2.048\times10^{-4}$
cm$^{3}$/mole based on the vanadium-based $S=1/2$ spinel compound
LiV$_{2}$O$_{4}$.\cite{Van-Vleck LiV2O4} From the Curie constant
obtained from the fit (0.30 cm$^{3}$K/mole V$^{4+}$), the effective
number of Bohr magnetons is found to be $1.55$ which is slightly
smaller than the value for a $S=1/2$ system ($\lyxmathsym{\textgreek{m}}_{\text{eff}}=1.73\lyxmathsym{\textgreek{m}}_{\text{B}}$).
Whereas no sharp anomaly is seen in the susceptibility data down to
2 K, a difference betwen the ZFC and FC susceptibilities is seen below
about 6 K suggestive of freezing of moments or glassy behaviour. Such
ZFC/FC bifurcation has been seen in Zn or Ti doped LiV$_{2}$O$_{4}$
around the same temperature. The asymptotic Curie-Weiss temperature
is large and negative ($\theta_{CW}$ $\simeq-214$ K), in comparison
to the freezing temperature, indicating strong antiferromagnetic interactions.
Note that the peak of the ZFC curve (at $T\sim3.5$ K) is less than
the temperature where the ZFC/FC bifurcation starts. The ZFC/FC split
is possibly linked with the onset of spin-cluster formation and the
peak of the ZFC curve indicates the freezing temperature ($T_{f}$).
These anomalies point to the fact that the freezing phenomena might
set in already above $T_{f}$. Similar kind of behavior has also been
observed in Li$_{1-x}$Zn$_{x}$V$_{2}$O$_{4}$ (for $x$$\geq0.8$\cite{magnetic and structural transition in LixZn1-xV2O4})
and in Li(V$_{1-y}$Ti$_{y}$)$_{2}$O$_{4}$ (for $y=0.2$\cite{Ground state properties od LiV2O4 and Li1-xZnx(V1-yTiy)2O4})The
inherent geometric frustration of the spinel system coupled with the
random occupation of the B sites by magnetic V$^{4+}$ and nonmagnetic
Li$^{+}$ is suggested to be responsible for the observed susceptibility
behaviour. Note that in comparison, the $\theta_{CW}$ values for
the Li(V$_{1-y}$Ti$_{y}$)$_{2}$O$_{4}$ ($y=0.05-1$) system are
less than 100 K. \cite{Ground state properties od LiV2O4 and Li1-xZnx(V1-yTiy)2O4}.
In our ac susceptibility measurements as well, a hump-like anomaly
was observed at 4 K in the real part of the susceptibility $\chi'$
(see Fig.\ref{fig:4ac susceptibility}). In the range of frequencies
of the ac field considered by us (20-1000 Hz), no shift in the peak
position of $\chi'$ was observed. This is different from the conventional
spin-glass behavior observed in the Zn-doped (at A-site) of Li$_{1-x}$Zn$_{x}$V$_{2}$O$_{4}$
\cite{Spin-glass behavior in Li1-xZnxV2O4}. This suggests that there
exists some kind of spin freezing in this compound but it is not the
conventional spin glass behaviour. 

\begin{figure}
\includegraphics[scale=0.34]{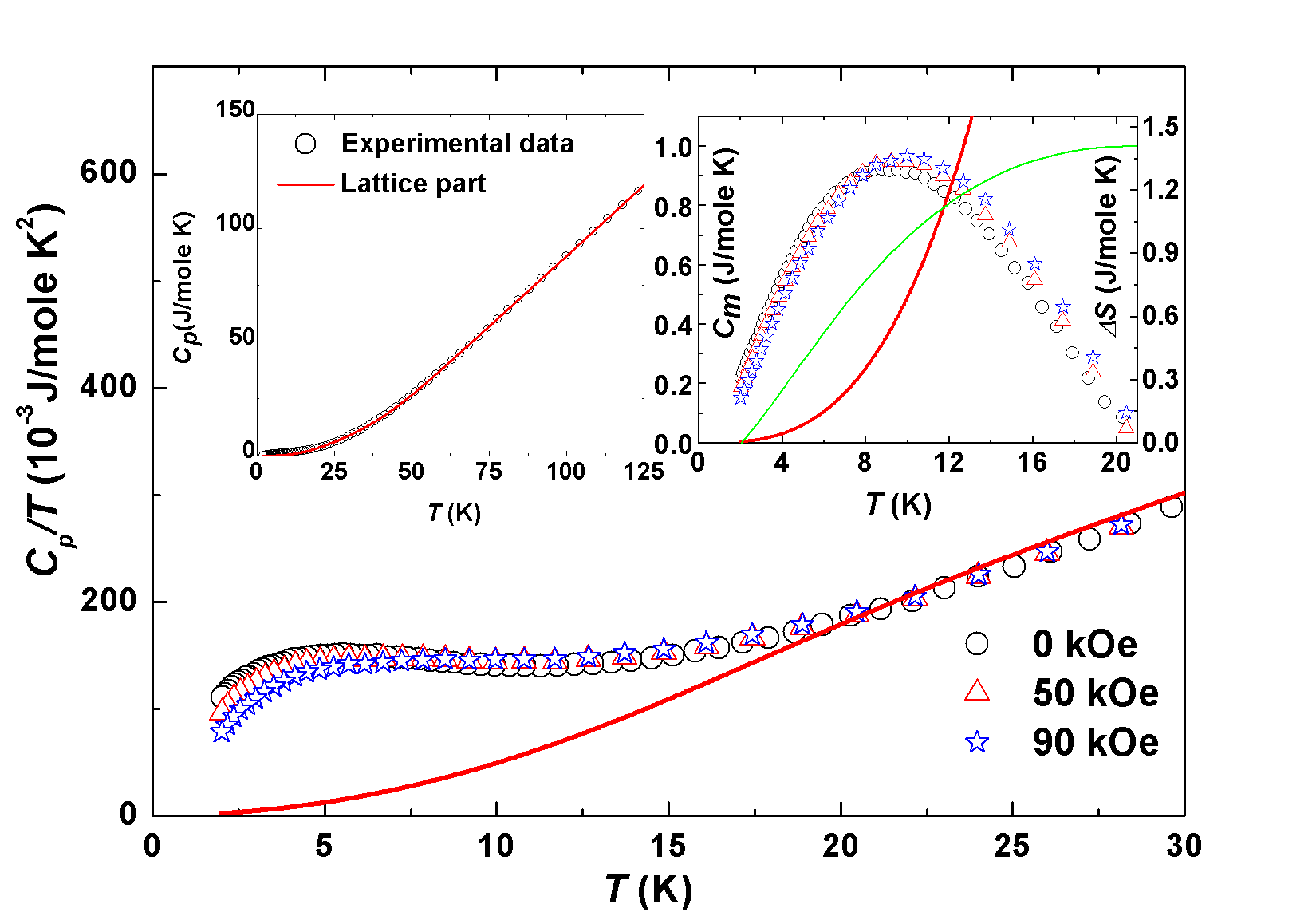}

\caption{\label{fig:5heat capacity}(Left inset) The temperature dependence
of specific heat of Li$_{2}$ZnV$_{3}$O$_{8}$; the red line represents
the fitting of the heat capacity using one Debye and two Einstein
terms (see text). The right inset displays the magnetic contribution
of specific heat at zero field (black circle), 50 kOe (red triangle),
and 90 kOe (blue star), the red line indicates the lattice contribution
of heat capacity; the green line (right axis, left inset) shows the
change of entropy calculated by integrating the $C_{\text{m}}/T$
data. In the main figure $C_{P}/T$ vs. $T$ is shown to depict the
slight change with the magnetic field. The red line indicates the
lattice contribution in $C_{P}/T$. Note that in both the insets and
in the main figure the heat capacity is calculated for one formula
unit. }
\end{figure}

The heat capacity data of Li$_{2}$ZnV$_{3}$O$_{8}$ are shown in
Fig.\ref{fig:5heat capacity}. It is clear from the figure that no
sign of long-range ordering is observed down to 2 K. At around 5 K
a small hump is seen in the $C_{\text{p}}/T$ data and it changes
slightly with the change in the magnetic field however this change
with magnetic field is not due to the Schottky effect. This hump in
the heat capacity data is close to the temperature where ZFC and FC
curves are split in the $\chi(T)$ vs. $T$ data. Since we do not
have a suitable non-magnetic analogue for this system we tried to
extract the magnetic specific heat of Li$_{2}$ZnV$_{3}$O$_{8}$
by subtracting the lattice contribution using a combination of Debye
and Einstein heat capacities, $C_{Debye}$ and $C_{Einstein}$, respectively.
In the $T$-range 27-125 K, the measured heat capacity $C_{P}$ could
be fit with a combination of one Debye and two Einstein functions
of the type given below where the coefficient $C_{d}$ stands for
the relative weight of the acoustic modes of vibration and the coefficients
$C_{e_{1}}$ and $C_{e_{2}}$ are the relative weights of the optical
modes of vibration. 

$C_{Debye}=\ensuremath{C_{d}\times9nR}\ensuremath{(T/\theta_{\text{d}})^{\mbox{\text{3}}}}\ensuremath{\int_{\text{0}}^{\text{\ensuremath{\theta\ensuremath{_{\text{d}}}/T}}}(x^{\text{4}}e^{\text{\ensuremath{x}}}/(e^{\text{\ensuremath{x}}}-1)^{\text{2}}}\ensuremath{)dx}$

$C_{Einstein}=\ensuremath{3nR[\sum C{}_{e_{m}}\times\frac{x_{E_{m}}^{2}e^{x_{E_{m}}}}{(e^{x_{E_{m}}}-1)^{2}}]}$,
$x=\frac{h\omega_{E}}{k_{B}T}$

In the above formula, $n$ is the number of atoms in the primitive
cell, $k_{\text{B}}$ is the Boltzmann constant, and $\theta_{\text{d}}$
is the relevant Debye temperature, $m$ is an index for an optical
mode of vibration. In the Debye-Einstein model the total number of
modes of vibration (acoustic plus optical) is equal to the total number
of atoms in the primitive unit cell. For Li$_{2}$ZnV$_{3}$O$_{8}$
this number is $14$. In this model we have considered the ratio of
the relative weights of acoustic modes and sum of the different optical
modes to be $1:n-1$. Due to having two light atoms (lithium, oxygen)
and two comparatively heavier atoms (vanadium and zinc) in this compound
we considered two different optical modes of vibrations. The fit yields
a Debye temperature of $157$ K and Einstein tempertures of $293$
K and $698$ K with relative weights $C_{d}:C_{e_{1}}:C_{e_{2}}$
$=$$1:4.3:8.7$. Upon subtracting the lattice heat capacity with
the above parameters, we obtain the magnetic contribution to the heat
capacity $C_{\mathrm{m}}(T)$. The entropy change ($\Delta S)$ was
calculated by integrating the $C_{m}/T$ data {[}see Fig. \ref{fig:5heat capacity}{]}.
The entropy change from about 25 K to 2 K is about 1.4 J/K ( calculated
for one formula unit containing 3 V$^{4+}$ ions) which is only about
8\% of the value for an $S=1/2$ system ($Rln(2S+1))$ which indicates
the presence of many degenerate low-energy states at low temperatures\cite{note entropy error}.
The value of the entropy change in our system is within the range
of values inferred for other doped LiV$_{2}$O$_{4}$ systems\cite{Ground state properties od LiV2O4 and Li1-xZnx(V1-yTiy)2O4}.
This large reduction in the value of $\Delta S$ down to temperatures
much lower than the Weiss temperature ($\theta_{CW}$) is typical
of disordered systems and a consequence of the presence of strong
geometric frustration in Li$_{2}$ZnV$_{3}$O$_{8}$. We observed
a broad maximum in the $C_{\mathrm{m}}(T)$ vs. $T$ data at 9 K.
Although we have not observed any frequency dependence in the $ac$
susceptibility data which is the main characteristic feature of canonical
spin-glasses but the maximum of $C_{\mathrm{m}}(T)$ is observed above
the freezing temperature ($T_{f}$). At low $T$, (in the range of
$2-5$ K) $C_{\mathrm{m}}(T)$ follows power law($C_{\mathrm{m}}(T)$
= $\gamma T^{\alpha}$with $\gamma=98.4$ mJK$^{-2.2}$mol$^{-1}$
and $\alpha$= $1.24$) dependence with temperature. This is similar
to what has been observed in other strongly correlated spin-glasses\cite{Ground state properties od LiV2O4 and Li1-xZnx(V1-yTiy)2O4,power law Cm URh2G2}.
All these facts tie together to point out that there might be some
formation of cluster-glass like state in the system around 9 K which
ultimately drives the system in the metastable frozen state below
$T_{f}$.

We were unable to detect the NMR signal associated with the $^{51}$V
nucleus of Li$_{2}$ZnV$_{3}$O$_{8}$. This could be due to the fact
that there is a strong, on-site local moment which naturally couples
well with its own nucleus. The fluctuations of this moment are very
effective in causing a fast relaxation of the nuclear magnetization.
This makes the detection of its NMR signal difficult. In Cs$_{2}$CuCl$_{4}$\cite{M.A.Vachon New journal of physics-8(2006)-133Cs-NMR-Cs2CuCL4}
and in BaV$_{3}$O$_{8}$\cite{BaV3O8} as well, an NMR signals from
the $^{63,65}$Cu and from magnetic V$^{4+}$nuclei, respectively
were not detected probably for similar reasons. On the other hand,
we did not face any difficulty in observing the NMR signal from the
$^{7}$Li nucleus in Li$_{2}$ZnV$_{3}$O$_{8}$. However, we did
not observe any temperature dependence of shift in the $^{7}$Li spectra
down to 6 K. From 6 K to 4 K (temperature close to the freezing temperature)
a small shift has been observed in the $^{7}$Li spectra. We point
out that in all the LiV$_{2}$O$_{4}$ based spinels reported in literature,
a significant temperature dependence of the shift (scaling with the
spin susceptibility) was observed. A weak hyperfine coupling may be
a reason for not getting any temperature dependence of shift in Li$_{2}$ZnV$_{3}$O$_{8}$
though that would be unusual since it has the same structure as LiV$_{2}$O$_{4}$
and one expects to have at least half the Li at the A site.

\begin{figure}
\includegraphics[scale=0.34]{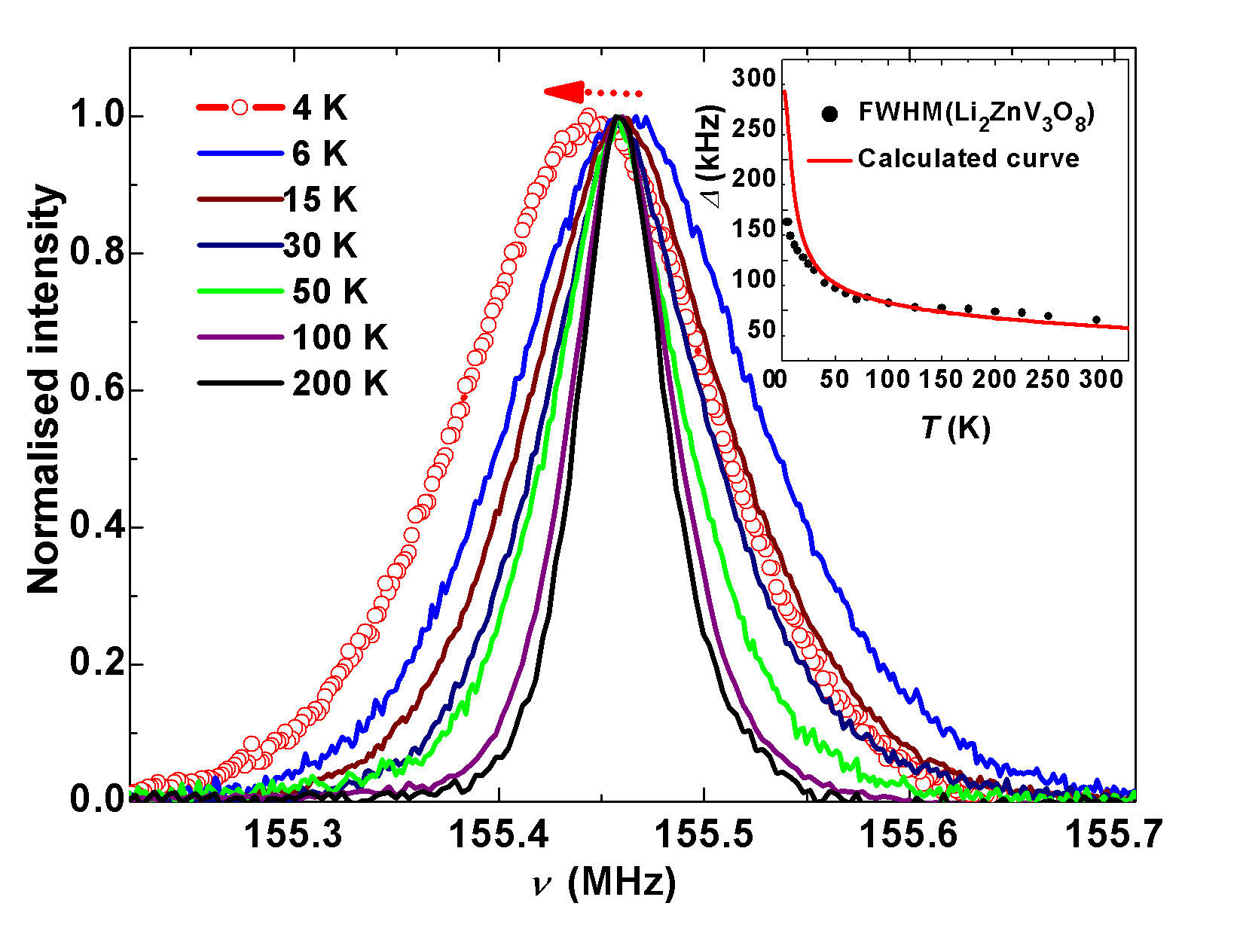}

\caption{\label{fig:6 spectra FWHM}The $^{7}$Li spectra of Li$_{2}$ZnV$_{3}$O$_{8}$
at temperatures down to 4 K. The dotted arrow points to the shift
from 6 K to 4 K. The inset shows the change in the FWHM (full width
at half-maximum) with the variation in $T$. The red solid line in
the inset indicates the calculated results discussed in the text.}
\end{figure}

The $^{7}$Li NMR linewidth increases with decreasing temperature
as shown in the inset of Fig. \ref{fig:6 spectra FWHM}. Down to about
100 K, the full-width-at-half-maximum (FWHM) of the $^{7}$Li spectrum
for Li$_{2}$ZnV$_{3}$O$_{8}$ is similar to that for pristine LiV$_{2}$O$_{4}$
and also not much different from the other doped-LiV$_{2}$O$_{4}$
systems. With a decrease in temperature the spectrum of Li$_{2}$ZnV$_{3}$O$_{8}$
broadens at a significantly slower rate compared to LiV$_{2}$O$_{4}$
and the doped-LiV$_{2}$O$_{4}$ systems\cite{Li NMR studies of Li1-xZnxV2O4 and Li(V1-yTiy)2O4}.
We now discuss the origin of $^{7}$Li NMR linewidth in Li$_{2}$ZnV$_{3}$O$_{8}$.
There are two contributions to the NMR linewidth. The first one originates
from the nuclear-nuclear dipolar interaction while the second arises
from the demagnetizing field due to the neighbouring powder grains\cite{51V linewidth and volume suscepyibilityMahajan}.
By using the Gaussian approximation we can write the $^{7}$Li linewidth
(FWHM) as

\begin{equation}
\Delta=2.35\sqrt{\text{\ensuremath{\lyxmathsym{\guilsinglleft}}\ensuremath{\Delta\nu}\ensuremath{\lyxmathsym{\guilsinglright}^{2}}}+(B\chi_{V}H\gamma/2\pi)^{2}}\label{eq:width}
\end{equation}

where $\ensuremath{\text{\guilsinglleft}}\ensuremath{\Delta\nu}\ensuremath{\text{\guilsinglright}}$
is the dipolar interaction term calculated from the internuclear interaction,
$\chi_{V}$ is the volume susceptibility ($\chi_{V}$ = $\chi_{M}$$d/M$,
$d$ = 4.134 g/cm$^{3}$ is the density, and $M$ = 360.09 g/mol is
the molar mass), $\gamma/2\pi$=1654.6 Hz G$^{-1}$, $H$ = 93954.3
Oe and $B$ is the fractional root-mean-square deviation of the local
field from the applied field $H$. The temperature independent dipolar
contribution has been calculated for LiV$_{2}$O$_{4}$ in Ref. \cite{7Li linewidth and volume suscepyibility Imai}
and is sure to be smaller in Li$_{2}$ZnV$_{3}$O$_{8}$ since it
is vanadium deficient and Zn does not have a nuclear moment. In the
first approximation, we take the dipolar width to be the same as in
LiV$_{2}$O$_{4}$ and fit the FWHM of Li$_{2}$ZnV$_{3}$O$_{8}$
to equation \ref{eq:width} with $B$ as a fitting parameter. For
$B=9.8$ the calculated curve agrees well with the experimental data
above 30 K (see Fig.\ref{fig:6 spectra FWHM}). At lower temperatures
the discrepancy between the experimental data and the calculated curve
might be due to (a) an extrinsic paramagnetic contribution to the
measured susceptibility which becomes larger at lower temperatures
and (b) an insufficient spectral width of the NMR pulses larger linewidth
at low-temperatures leading to an underestimation of the experimental
FWHM. The overall behaviour is similar to that seen in Ref. \cite{7Li linewidth and volume suscepyibility Imai}
and hence we conclude that here as well, the $^{7}$Li FWHM is dominated
by macroscopic magnetisation effects. 

We next report the $^{7}$Li spin-lattice relaxation rate data. In
Li$_{2}$ZnV$_{3}$O$_{8}$, lithium ($I=3/2)$ is in the tetrahedral
(A-site) and octahedral (B-site) environment with oxygen. At least
at the tetrahedral sites (but also at the octahedral sites in case
of distortions) a non-zero electric field gradient (EFG) quadrupolar
splitting is expected. However, down to 4 K, we were able to irradiate
the full spectrum. We have measured the recovery of the longitudinal
$^{7}$Li nuclear magnetization $M_{t}$ after a saturating comb and
fitted it using the following stretched exponential function,
\begin{equation}
1-M_{t}/M_{\infty}=A\times e^{-(t/T_{1})^{\beta}}
\end{equation}

\begin{figure}
\includegraphics[scale=0.33]{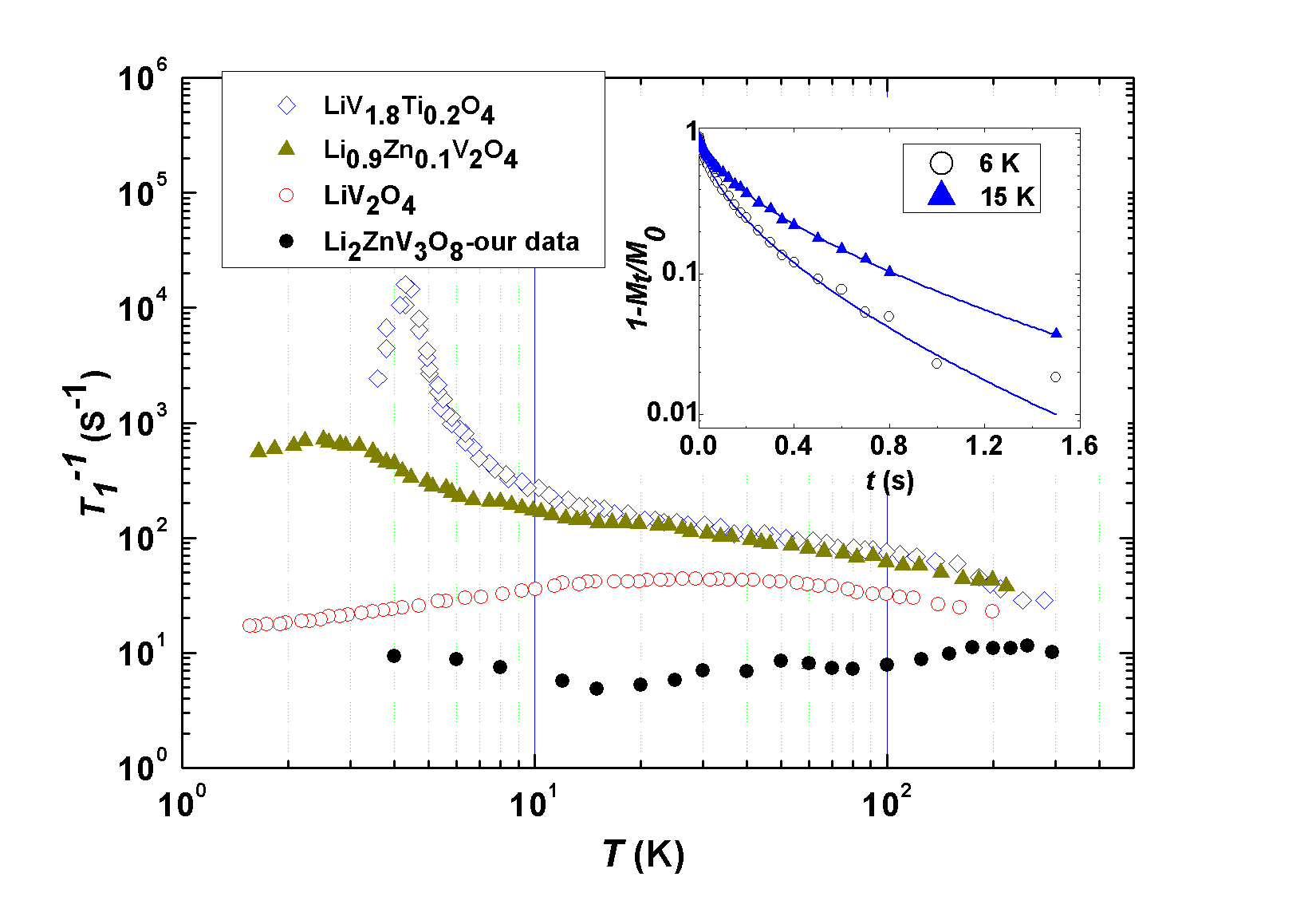}\caption{\label{fig:7 1/T1 comparisons}The spin-lattice relaxation rate (1/$T_{1}$)
of $^{7}$Li in Li$_{2}$ZnV$_{3}$O$_{8}$ is compared to that in
pure and doped LiV$_{2}$O$_{4}$ compounds from literature. In this
graph Li$_{2}$ZnV$_{3}$O$_{8}$, LiV$_{2}$O$_{4}$, Li$_{1-x}$Zn$_{x}$V$_{2}$O$_{4}$($x=0.1$)
and Li(V$_{1-y}$Ti$_{y}$)$_{2}$O$_{4}$($y=0.1$) data are shown
in black closed circle, red open circle, dark yellow closed triangle
and blue open diamond symbols, respectively. In the right inset the
saturation recovery data for $^{7}$Li nuclear magnetisation are shown
at 6 K and 15 K are shown. Also shown are the stretched exponential
fits by the blue solid lines. }
\end{figure}

\begin{figure}
\includegraphics[scale=0.35]{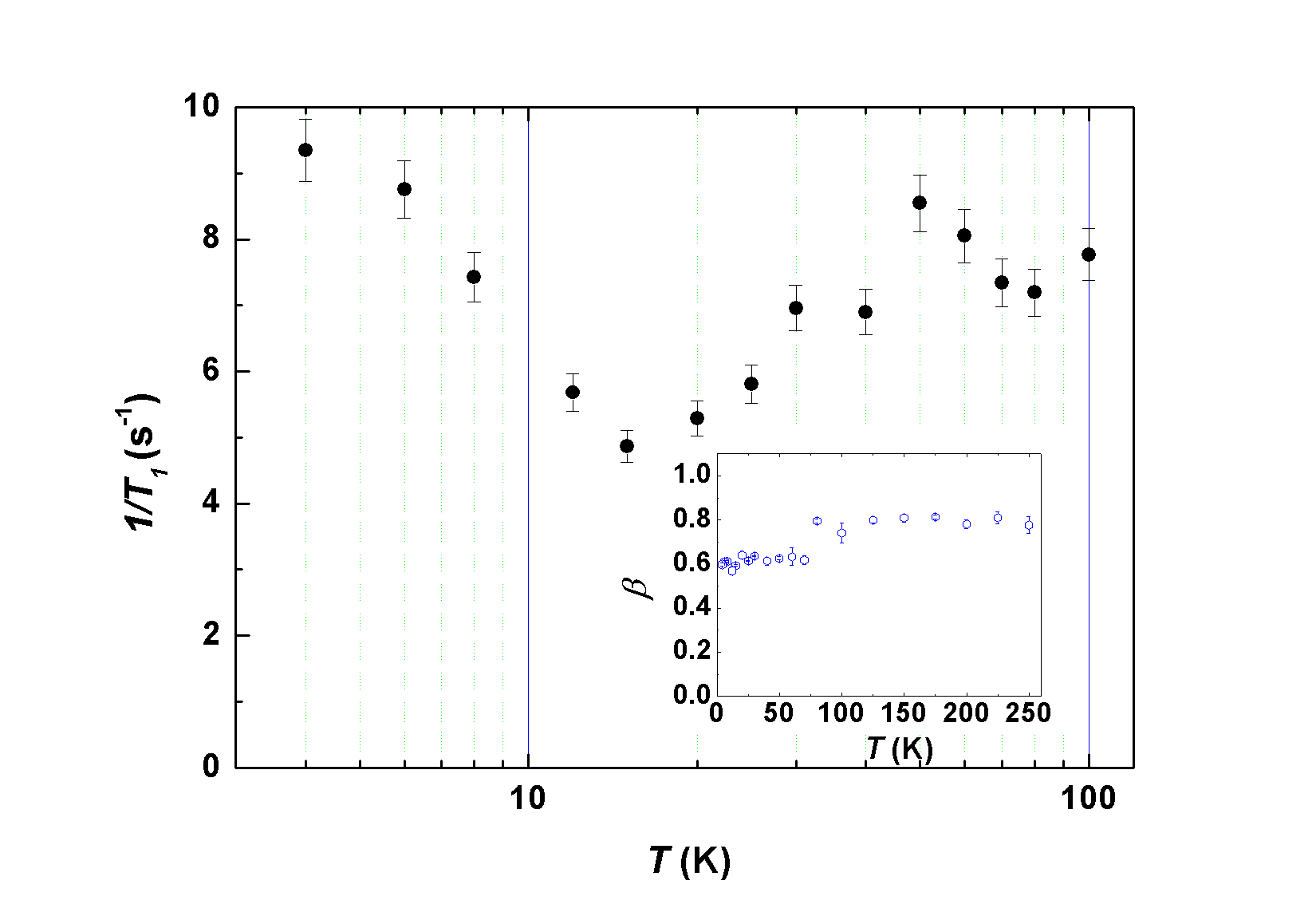}

\caption{\label{fig:8 1/T1 exponents}The temperature dependence of the $^{7}$Li
spin-lattice relaxation rate ($1/T_{1}$) is shown. The inset shows
the variation of the stretching exponent $\beta$ with temperature.}

\end{figure}

Here $A$ stands for the amount of saturation and $\beta$ denotes
the stretching exponent. The systems which are driven into a spin-glass-like
state possess a distribution of spin-lattice relaxation times due
to different relaxation channels with different $T_{1}$$.$ In these
situations $\beta$ is a measure for the width of the distribution.
This stretched exponential behavior of the saturation recovery of
the spin-lattice relaxation data gives an indication of the presence
of the local moments, although sensed very weakly through the window
of the lithium nucleus. The $^{7}$Li NMR $1/T_{1}$ is nearly unchanged
with temperature \cite{note} which is again surprising in light of
the published data on pure and doped LiV$_{2}$O$_{4}$ (see Fig.
\ref{fig:7 1/T1 comparisons} where literature data are shown along
with our data). We observed an increase in our $^{7}$Li NMR $1/T_{1}$
data for Li$_{2}$ZnV$_{3}$O$_{8}$ near the spin-glass/freezing
temperature (4 K) and a small hump-like anomaly around 50 K (see Fig.\ref{fig:8 1/T1 exponents}).
Likewise in case of Li$_{1-x}$Zn$_{x}$V$_{2}$O$_{4}$ and Li(V$_{1-y}$Ti$_{y}$)$_{2}$O$_{4}$,
an anomaly/peak in the $T$-dependence of $^{7}$Li NMR $1/T_{1}$
was seen near the spin-glass/freezing temperature. Note that the typical
value of $^{7}$Li NMR $1/T_{1}$ data in Li$_{2}$ZnV$_{3}$O$_{8}$
($\sim8s^{-1}$) is vastly smaller than in any of the other vanadium
based spinels stated above. In $^{7}$Li NMR, the only indication
of the presence of local moments in Li$_{2}$ZnV$_{3}$O$_{8}$ is
in the temperature dependence of the linewidth and in the stretched
exponential behavior in the spin-lattice relaxation. 

Very recently $^{7}$Li NMR studies of a newly-found valence-bond
geometrically-frustrated ($S=1/$2) cluster magnet system (LiZn$_{2}$Mo$_{3}$O$_{8}$)
with strong correlation ($\theta_{CW}\simeq-220$ K) between Mo$_{3}$O$_{13}$
clusters via oxygen bridges has been reported where no shift has been
observed down to 4.2 K and an almost temperature-independent spin-lattice
relaxation rate ($1/T_{1}$) of the order of $4s^{-1}$has been observed
for the main peak\cite{LiZn2Mo3O8} similar to what we have seen in
Li$_{2}$ZnV$_{3}$O$_{8}$.

\section{Conclusion}

In this work we have reported a new vanadium-based cubic spinel (AB$_{2}$O$_{4}$)
system Li$_{2}$ZnV$_{3}$O$_{8}$, its preparation, crystal structure,
magnetic properties, specific heat properties, and NMR measurements.
Among the B-site doped LiV$_{2}$O$_{4}$ systems, this is possibly
the first compound where all the vanadium ions are in the 4$^{+}$
oxidation state. Whereas the B-sites form a frustrated lattice only
75\% are occupied by V$^{4+}$ and the others have non-magnetic ions.
A Curie-Weiss fit of the magnetic susceptibility data yields a Curie
constant close to that expected for $S=1/2$. The Curie-Weiss temperature$\theta_{CW}$=
-214 K (greater than in the other LiV$_{2}$O$_{4}$ variants) is
suggestive of antiferromagnetic correlation in the system. Spin freezing
is observed below 6 K, similar to that in other LiV$_{2}$O$_{4}$
variants. The peak of the ZFC curve ($T\sim$3.5 K) appears lower
than the bifurcation point of the ZFC/FC curves ($T\sim6$ K) indicating
that with decreasing temperature there might be an onset of spin-cluster
formation before the system is ultimately driven into the frozen state.
In the heat capacity measurement an anomaly is seen around 9 K and
the entropy change $\Delta S$ is only 8\% of that expected for an
ordered $S=1/2$ system. This is likely due to the presence of strong
geometric frustration in the system. We were unable to detect the
NMR signal associated with the $^{51}$V nucleus of Li$_{2}$ZnV$_{3}$O$_{8}$
due to strong on-site local moment so we worked with the $^{7}$Li
nucleus. No temperature dependence of $^{7}$Li NMR shift (except
near the freezing temperature of 4 K) was observed which might indicate
a weak hyperfine coupling with the magnetic V$^{4+}$. The saturation
recovery of the spin-lattice relaxation data has been fitted well
using a stretched exponential function as might happen with a distribution
of magnetic environments. In the temperature dependence of the spin-lattice
relaxation rate we observed an increase close to the freezing temperature
(4 K) but the magnitude of the anomaly close to the freezing temperature
is 2-3 orders of magnitude smaller than for Li$_{1-x}$Zn$_{x}$V$_{2}$O$_{4}$
and Li(V$_{1-y}$Ti$_{y}$)$_{2}$O$_{4}$.

\section{Acknowledgement}

Discussions with R.K.Sharma are acknowledged. The authors thank the
Department of Science and Technology, Govt. of India for financial
support.


\begin{thebibliography}{References}
\bibitem[18]{note entropy error} Depending on the temperature range
chosen for the fit, the inferred lattice heat capacity is a bit different.
Also, since this system has a large $\theta_{CW}$ (= -214 K), entropy
release might take place in a broad temperature range below about
200 K. Due to the large lattice contribution at high temperatures
it is nearly impossible to accurately determine the magnetic contribution
to the heat capacity in this range. Consequently, there is likely
to be a significant uncertainty in the entropy change estimation.
\bibitem[24]{note}The small change in $\beta$ seen around 80 K might
be an artifact related to the changing of cryogen from nitrogen to
helium and the need to use lower $rf$ power at lower temperatures.
\end{thebibliography}
\end{document}